\documentclass[12pt, letterpaper]{article}
\usepackage[utf8]{inputenc}
\usepackage[english]{babel}
\usepackage{authblk}
\usepackage{shortcuts}
\usepackage{amsmath}
\usepackage{float}
\usepackage{txfonts,graphicx}
\usepackage{makecell}

\usepackage[hyphens]{url}
\usepackage[hyphenbreaks]{breakurl}

\usepackage{natbib}
\title{Blocks as geographic discontinuities:\\ The effect of polling place assignment on voting}

\author[1]{Sabina Tomkins}
\author[1]{Keniel Yao}
\author[1]{Johann Gaebler}
\author[2]{Tobias Konitzer}
\author[2,3]{David Rothschild}
\author[4]{Marc Meredith}
\author[1]{Sharad Goel}

\affil[1]{Stanford University}
\affil[2]{PredictWise}
\affil[3]{Microsoft Research}
\affil[4]{University of Pennsylvania}

\proofreadfalse
\showremovalsuggestionsfalse
\graphicspath{{figures/}}

\begin{document}

\maketitle

\begin{abstract}
A potential voter must incur a number of costs in order to successfully cast an in-person ballot, including the costs associated with identifying and traveling to a polling place. In order to investigate how these costs affect voting behavior, we introduce two quasi-experimental designs that can be used to study how the political participation of registered voters is affected by differences in the relative distance that registrants must travel to their assigned Election Day polling place and whether their polling place remains at the same location as in a previous election. Our designs make comparisons of registrants who live on the same residential block, but are assigned to vote at different polling places. We find that living farther from a polling place and being assigned to a new polling place reduce in-person Election Day voting, but that registrants largely offset for this by casting more early in-person and mail ballots.

\end{abstract}

\section{Introduction}

It has long been accepted by political scientists that the choice by a potential voter to vote or abstain from voting is based on a assessment of whether the benefits from voting are higher than the costs \citep{RikerOrdeshook1968}. The opportunity cost of the time spent casting a ballot is one of the most important costs that a potential voter must incur in order to vote \citep{Downs1957}. There are concerns that even small increases in expectations about the time it takes to vote could reduce turnout, particularly given that many potential voters receive limited expected benefits from voting \citep{Aldrich1993}.

Potential voters are likely to consider the search costs and transportation costs associated with casting an in-person ballot \citep{BradyMcNulty2011}. Search costs refer to the cost of identifying where a polling place is located and how to get there, and are thought to decrease when a potential voter repeatedly votes at the same polling place. Transportation costs refer to the cost of traveling to a polling place, and will increase, typically, as a polling place moves further from a potential voter's residence.

Political science research shows that turnout decisions are affected by changes in search and transportation costs. \citet{BradyMcNulty2011} find that potential voters are two percentage points less likely to cast an in-person ballot on Election Day when they were assigned to vote at a new polling place that was equally far from their house. Two percentage points represents the median estimated reduction in in-person voting on Election Day from a polling place change in existing work, with \citet{McNultyDowlingAriotti2009} and \citet{AmosSmithSteClaire2017} finding more than a two-percentage point decline, and  \citet{yoder2018polling} and \citet{clinton2019polling} finding less. \citet{cantoni2020precinct} also shows that potential voters who live in the same neighborhood are less likely to vote when the polling place that they are assigned to vote at on Election Day is further from their residence.

Existing research differs in whether increasing the search and transportation costs associated with Election Day voting primarily causes potential voters to \textit{abstain} or \textit{substitute} to early in-person voting or mail balloting. \citet{clinton2019polling} show that most potential voters dissuaded from voting in-person on Election Day by increases in search and transportation costs switched to early in-person voting.  In contrast, \citet{BradyMcNulty2011} find that about 60 percent of the potential voters who were dissuaded from voting in-person on Election Day because of higher search costs abstained, with the other 40 percent of shifting to mail ballots. Likewise, \citet{AmosSmithSteClaire2017} find that about 60 percent of the potential voters who were dissuaded from voting in-person on Election Day because of higher search costs abstained, with the other 40 percent of shifting to early in-person voting or mail balloting.

Synthesizing the existing evidence is challenging because studies vary both in the context that is being studied and the specific design being applied to conduct the study. States vary in the ease with which voters can substitute into using mail ballots or early in-person voting as the cost of voting in person increases. Some designs focus on the impact of transportation costs among those who are experiencing an increase in search costs, while others focus only on cases in which there are no changes in polling locations. Designs also differ in whether they focus exclusively on polling place changes resulting from the consolidation of polling places, or consider the larger set of polling place changes caused by consolidation, expansion, and movement of polling places. These differences make it hard to infer how the results of a given study may generalize to alternative contexts in which substitutes are more or less accessible or there is a different cause of the differences in voting costs. 

We introduce two new quasi-experimental designs to study the consequences of increased search and transit costs on potential voters in a wide variety of contexts. In brief, our designs leverage cases in which registrants who live on one side of a street are assigned to vote at a different polling place on Election Day than registrants who live on the other side of the same street. This allows us to make comparisons of otherwise similarly situated registrants who travel different distances to their polling place or differ in whether they are assigned to vote at a new polling place or not, and ascribe differences in their turnout patterns to the differential costs associated with finding and getting to their assigned polling place. Our primary analysis focuses on data from ten states  in which we were able to: (1) collect information of polling place addresses in the 2016 and 2012 presidential elections; (2) link these data with voter registration records; and 
(3) discern from the voter
 registration record the mode in which a 2016 vote was cast (i.e., mail-ballot, early in-person ballot, or polling place). We also perform some additional analyses for the six of these ten states in which there were sufficient observations to estimate the effect of being assigned to a new polling place. 
 
Our designs have a number of features that make it relatively easy to apply to many states simultaneously in order to generate the most comprehensive analysis to date of registrants' ability to offset voting costs with substitutes. First, these designs incorporate data from many types of jurisdictions. Our approach for studying relative distance uses information from any jurisdiction in which registrants on the same block are assigned to a different polling location, which occurs frequently. Similarly, our approach for studying shocks considers polling place changes that result from the consolidation, the expansion, and the movement of polling places. Second, residential address, the key variable that we exploit to generate variation in the cost of in-person voting, is a variable that is available in any voter registration database. Third, our approach does not require states to provide any information beyond the mapping between precincts and polling locations and the addresses of the polling locations. Importantly, this means we are not limited to only studying states that make it easy to access the mostly non-existent data files of precinct boundaries.

Our results show that the ability to substitute into other modes of voting is key to helping mitigate consequences of increased search and transit costs on Election Day voting. We find that a \unit{} who lives further from their polling places is approximately 1.6 percentage points less likely to vote at this polling place than a similarly situated \unit{} who lives closer, but offsets for this with their increased use of mail balloting or early in-person voting. Likewise, a \unit{} assigned to a different polling place in 2016 than in 2012 is about 1 percentage point less likely to vote in-person on Election Day than a similarly situated \unit{} who was assigned to the same polling place in both elections, but partially compensates for this by substituting into mail balloting or early in-person voting. These findings highlight the importance of making alternative voting methods available for potential voters who find it difficult to vote at their polling place on Election Day.

\section{Data}
\label{sec:data}

Our analysis uses state-supplied polling place files, which provide identifiers and addresses for polling places, and a national voter file, which provides \unit{} characteristics, voting behavior and voting place identifiers. Together these data allow us to infer which polling place  \units{} were assigned to vote at on Election Day and to compute the distance from their residences to this polling place. We then select blocks comprised of similar \units{} who are assigned to vote at different polling places and use variation in distance between the block and the polling places, as well as the stability of the location of polling places between two elections, to estimate the effects of relative distance and changes to polling places (henceforth shocks) on turnout decisions. The remainder of this section provides a sketch of the process of cleaning and combining our primary data sources, with more details provided in \secref{app:filtering}.
 
 To collect data on polling place locations, we filed public-records requests to state-election officials in all fifty states. The full account of these requests is included in \secref{app:states}. We received records from 18 states which met the basic criterion that they contain address and voting jurisdiction descriptors for both 2012 and 2016. A number of states, including Texas, reported that there was no statewide aggregation of polling place locations. A subset of 254 counties in Texas responded to our public record requests with data that met the basic criterion. Table \ref{table:example_places} shows a stylized example of the information these data contain. 
 
 We merge information on polling place locations to registration records in snapshots of TargetSmart's national voter file from November 2012 and November 2016. These data include a \unit's voting jurisdiction (i.e., precinct), registration address, and vote history. In some states, the vote history data also include information on which method of voting was used. 
 
Our design requires us to find blocks of \units{} in which some \units{} on the block are assigned to vote at one polling location on Election Day, while other \units{} on the same block are assigned to vote at a different polling location on Election Day. We create a block identifier for each address of registration that consists of all but the final two digits of the street number (e.g. 200 would be encoded as 2, and 2100 would be encoded as 21), the street name, the street type, and the city and state of a voter's residential address.
For example, the two \units{} in \tabref{table:example_voters}, residing at 123 Main St.\ and 125 Main St.\ in Milwaukee, Wisconsin, respectively, share
a block identifier. Table \ref{table:example_voters} shows a stylized example of these data after these block identifiers are generated. 

\begin{table}
    \begin{center}
    \begin{minipage}{.69\linewidth}
\resizebox{.99\columnwidth}{!}{%
\begin{tabular}{lllll}
\hline
\textbf{Place ID}&\textbf{Address}& \textbf{City} & \textbf{State} & \textbf{Precinct}  \\
\hline
\hline
PP-1 & 200 Main St & Milwaukee & WI & Cherry School 1\\
PP-2 & 1000 Third St & Milwaukee & WI & Apple School 1 \\
\hline
\end{tabular}
}
\caption{ Example polling place file. 
}\label{table:example_places}
 \end{minipage}%
 \end{center}
      \hfill\allowbreak%
 \begin{minipage}{.99\linewidth}
 \begin{center}
\resizebox{.99\columnwidth}{!}{%
\begin{tabular}{llllll}
\hline
\textbf{Voter ID}& \textbf{Address} & \textbf{City} &\textbf{State}& \textbf{Block ID}& \textbf{Precinct} \\
\hline
\hline
Voter-1 & 123 Main St& Milwaukee& WI& 1-Main-St-Milwaukee-WI &Cherry School 1 \\
Voter-2  & 125 Main St& Milwaukee& WI& 1-Main-St-Milwaukee-WI  &Cherry School 1  \\
Voter-3 & 2000 Third St& Milwaukee &WI& 20-Third-St-Milwaukee-WI &Apple School 1 \\
\hline
\end{tabular}

}
\caption{ Example voter records. 
}\label{table:example_voters}
\end{center}
   \end{minipage}%

\end{table}

To infer polling-place assignments we match \units{} to polling places with voting-jurisdiction metadata, such as precinct name and city. Consider the three example voters in \tabref{table:example_voters} and the two example polling places in \tabref{table:example_places}. Voter-1 and Voter-2 would be assigned to PP-1 while Voter-3 would be assigned to PP-2. \secref{app:filtering} discusses how we calculate distance between a \unit{}'s residential address and the assigned polling place once we construct these data.    

We apply filters to our data after matching \units{} to polling places to ensure that \units{} in the same block are similar except potentially for the location of their polling place. First, our analysis focuses on \units{} with the same registration address in the 2012 and 2016 snapshots to ensure changes in the location of polling places result from changes in polling place assignment rather than registrants moving. Second, we restrict our sample to blocks in which \units{} are assigned to two distinct polling locations. We also require that at least two \units{} were assigned to both of these polling locations. Third, we discard blocks in which there are two \units{} on that block that live more than \(0.3\)  miles\footnote{Small alterations to this threshold do not qualitatively change the results.} apart to ensure that a block constitutes a group of \units{} who live close to each other. Finally, many of our analyses restrict our sample to those states that provide information on voting method in their vote history.

\section{Methods: Measuring the effects of relative distance and shock}
Polling-place assignments can dictate many aspects of \units{}' voting experiences, including how long it takes to cast a ballot and how far one must travel to do so. We propose identification strategies that exploit the quasi-random assignment of \units{} living on the same block to different polling places to estimate two causal effects.

We refer to the first effect as the \textit{\distance}. Here we consider two simplified notions of distance when \units{} on a block are assigned to two polling places: farther and nearer. All \units{} on a block assigned to the nearer polling place are said to experience a nearer distance, while those on the face with a greater average distance are said to experience a farther distance. The \textit{\distance} is the difference in the likelihood that those \units{} who experience a farther distance vote using a given method relative to those \units{} who experience a nearer distance.   

We refer to the second effect as the \textit{\shock}. A shock occurs when a \unit{}, living at the same residence, is reassigned to a new polling place in between elections. The \textit{\shock} is the difference in the likelihood that those \units{} who experience a shock vote using a given method relative to comparable \units{} who were assigned to vote at the same polling place in both elections. Additionally, we require that one of two conditions holds: either that all \units{} were assigned to the same polling place in 2012, or, if \units{} were assigned to two different places that the average distance to each place be similar (that the difference in the average distance to each polling place be less than 0.25 miles). 

We are interested in the \distance{} and the \shock{} on turnout, as well as voting using specific methods. Thus, we examine the effect of both on all voting (\any), voting in person (\ip), and voting by a substitute method such as casting a mail-in or absentee ballot, or voting early in person (\sub).

 For each of the voting behaviors we examine, let
 
\[ votecast^\method_{\voter} =    
    \begin{cases}
      1, & \text{if voter } $\voter{}$  \text{ cast a vote by method} $\method$, \\
      0, & \text{otherwise}.
    \end{cases}
    \]
where $\method \in \{\text{\ip,\ \sub,\ \any} \}$.  
 We use a linear regression to estimate Equation \ref{eqn:effects} with standard errors clustered by household to account for autocorrelation in the unmodeled determinants of voting behavior by \units{} who reside in the same household.

 Let $\voter$ refer to an individual voter living in household $h$, $\treatmentset_\block$ be the set of all voters on block $\block$ assigned to treatment, $\controlset_\block$ be the set of all voters on block $\block$ assigned to control,  $\gamma_{b}$ be a block fixed effect, and $\treatmentvar_{\voter} = \mathbb{I}_{\voter \in \treatmentset_b}$ be an indicator of whether a registrant is assigned to treatment or control.

\begin{equation}
    \label{eqn:effects}
\votecast^\method_{\voter}  = \gamma_{b} + \theta\treatmentvar_{\voter}  + \epsilon_{i,h}
\end{equation}

We use \eqnref{eqn:effects} to estimate the corresponding \distance{} and \shock{}.

\subsection{Identification strategies}
\label{sec:shock_design}
\begin{figure}[t!]
\begin{center}
    \includegraphics[width=.99\columnwidth]{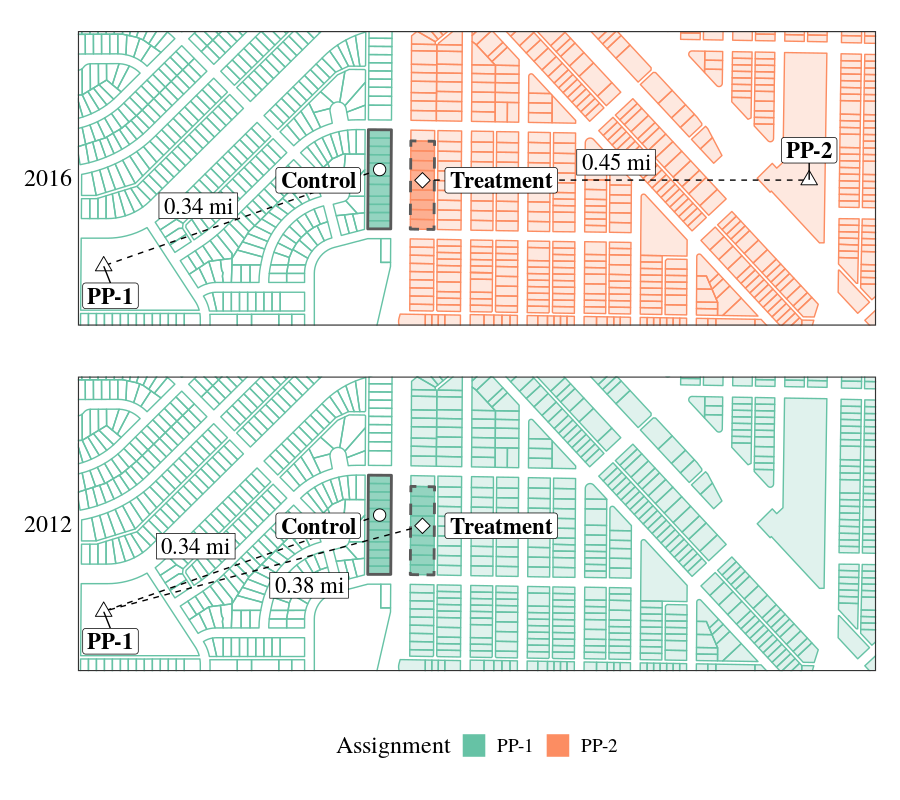}
    \caption{
    This map illustrates the block-randomization identification strategy for the \shock{} with two precincts in Milwaukee, WI across 2012 and 2016. Each precinct's color corresponds to its polling place assignment. In 2012 all \unit{}s in both precincts are assigned to PP-1. In 2016 all \unit{}s of the right-most precinct experience a shock as their polling place assignment changes from PP-1 to PP-2. To identify eligible blocks of voters for our analysis, we identify blocks on the boundaries of the two precincts;
    \unit{}s of the block face that experiences a shock (dashed outline) are assigned to treatment while \unit{}s of the block face that does not experience a shock (solid outline) are assigned to control. 
    \label{assignment-shock}}
    \end{center}
\end{figure}

To estimate the \distance{} we identify all blocks which lie on  the boundaries of two voting jurisdictions. For example, we show a block where one face (shown with a solid line) is assigned to a different polling place than the face across the street (shown with a dashed line) in the top panel of \figref{assignment-shock}.
 To compare the voting behavior of the \units{} who live farther from their polling place relative to those on the same block who live closer,
we require that a block contain \unit{}s assigned to two different polling places. Additionally, we require that these assignments remain unchanged between 2012 and 2016 as this allows us to isolate the effect of relative distance independent of \shock. For such blocks, \unit{}s assigned to the polling place with the lesser average distance to its \unit{}s are assigned to control, and \unit{}s assigned to the place with the greater average distance are assigned to treatment. This approach is similar to one used by \citet{MiddletonGreen2008}, which used variation in canvassing activity along the boundary of a precinct to estimate the effect of canvassing on turnout. 
The set of all relevant blocks defined by this identification strategy
 spans ten states (Hawaii,
Iowa,
Indiana,
Maryland,
North Carolina,
Pennsylvania,
Rhode Island,\
Texas,
Utah and
Wisconsin) and a total of 249,309 \units{}. 

Our identification strategy for the \shock{} is illustrated in \figref{assignment-shock}.
We locate all blocks which lie on  the boundaries of two precincts, such that a single face of this block experiences a polling-place shock between 2012 and 2016 while the other does not. 
\figref{assignment-shock} highlights two block faces outlined with dashed and solid borders. All \units{} who reside on the face enclosed with the dashed line experience a shock (and are assigned to treatment) and all those residing on the side enclosed with the solid line do not (and are assigned to control). 

Note that a shock may arise for multiple reasons. First, a reduction in the number of polling places, which is sometimes referred to as a consolidation, causes some registrants to be assigned to a new polling place that, on-average, will be located further away from their residence. Second, an increase in the number of polling places causes some registrants to be assigned to a new polling place that, on-average, will be located closer to their residence. Finally, a polling place being moved causes some registrants to be assigned to a new polling place without any clear expectation about how the change will affect the average distance between the polling location and the registrants' residence. Here, we do not differentiate between types of shocks; any block where one block face experiences a change and the other does not is included in our analysis regardless of whether the change is a result of a consolidation, addition or a movement. This contrasts our approach with some previous studies which focus specifically on the effects of consolidations \citep{BradyMcNulty2011, AmosSmithSteClaire2017}.

We use this identification strategy to select relevant blocks from the dataset constructed in \secref{sec:data}. In addition to the criteria detailed in \secref{app:filtering}, in the blocks used to estimate the \shock{} we require that each block face on the same block be a similar distance from its assigned polling place in 2012. In particular we require that the difference between the average distance to the polling place across the block faces be no more than 0.25 miles.
In total, the dataset used to estimate the \shock{} spans six states (Iowa, Indiana, Maryland, North Carolina, Pennsylvania and Wisconsin) and includes 47,456  \units.

Our identification strategies are examples of geographic discontinuity designs, which have been recently popularized in political science by \citet{KeeleTitiunik2015, KeeleTitiunik_2016}. Focusing only on comparisons within blocks will cause us to ignore many potential comparisons of \units{} who live in close proximity to one another, but are assigned to vote at different polling places. The benefit is that we expect there to be fewer differences in the underlying propensity to vote among \units{} on the same block than \units{} living on different blocks that are located in close proximity. And even when there are differences, there is no reason to expect that \units{} with a higher propensity to vote should systematically end up closer or further from their polling place than those with a lower propensity to vote. Thus, focusing on \units{} residing on the same block allows us to implement a geographic discontinuity design even when we know little about the specific \units{}, and when there are too many boundaries to apply a method like that proposed by \citeauthor{KeeleTitiunik2015} that empirically investigates whether housing prices are comparable on either side of a given boundary to determine whether we expect the underlying propensity to vote to be similar on either side of the boundary.

Our identification strategy assumes that registrants who live on different sides of a block would use voting methods at the same rate if not for differences in the polling places that registrants on each side of a block were assigned to vote at on Election Day. The figures that we present in Section \ref{sec:additional} of the Appendix examine how the characteristics of registrants assigned to different types of polling places differ. We first look at age, gender, modeled partisanship, and modeled race, as well as indicators representing that the total number of registrants on that side of the block fell into a certain range and find that all absolute differences are less than 2 percentage points and that the majority are less than 1 percentage point. For a subset of registrants for which we observe sale prices of homes,  we follow \citet{KeeleTitiunik2015} and examine whether the housing prices are comparable on the side of blocks that are closer and further from their polling place. We find that housing prices are similar for each side of the block. 

Similar patterns are found when we conduct the same analysis except defining those on the side of a block experiencing a polling place change as the treatment group and those of the side of a block keeping the same polling place as the control group. Unlike with our relative distance analysis, we also expect to observe similar voting patterns in the treatment and control groups in the previous election before the shock was realized. This is indeed the case, as we find similar historical voting patterns in both groups (please see Section \ref{sec:additional} of the Appendix).

\section{Results}

\begin{figure}[H]
\centering
    \includegraphics[width=.7\columnwidth]{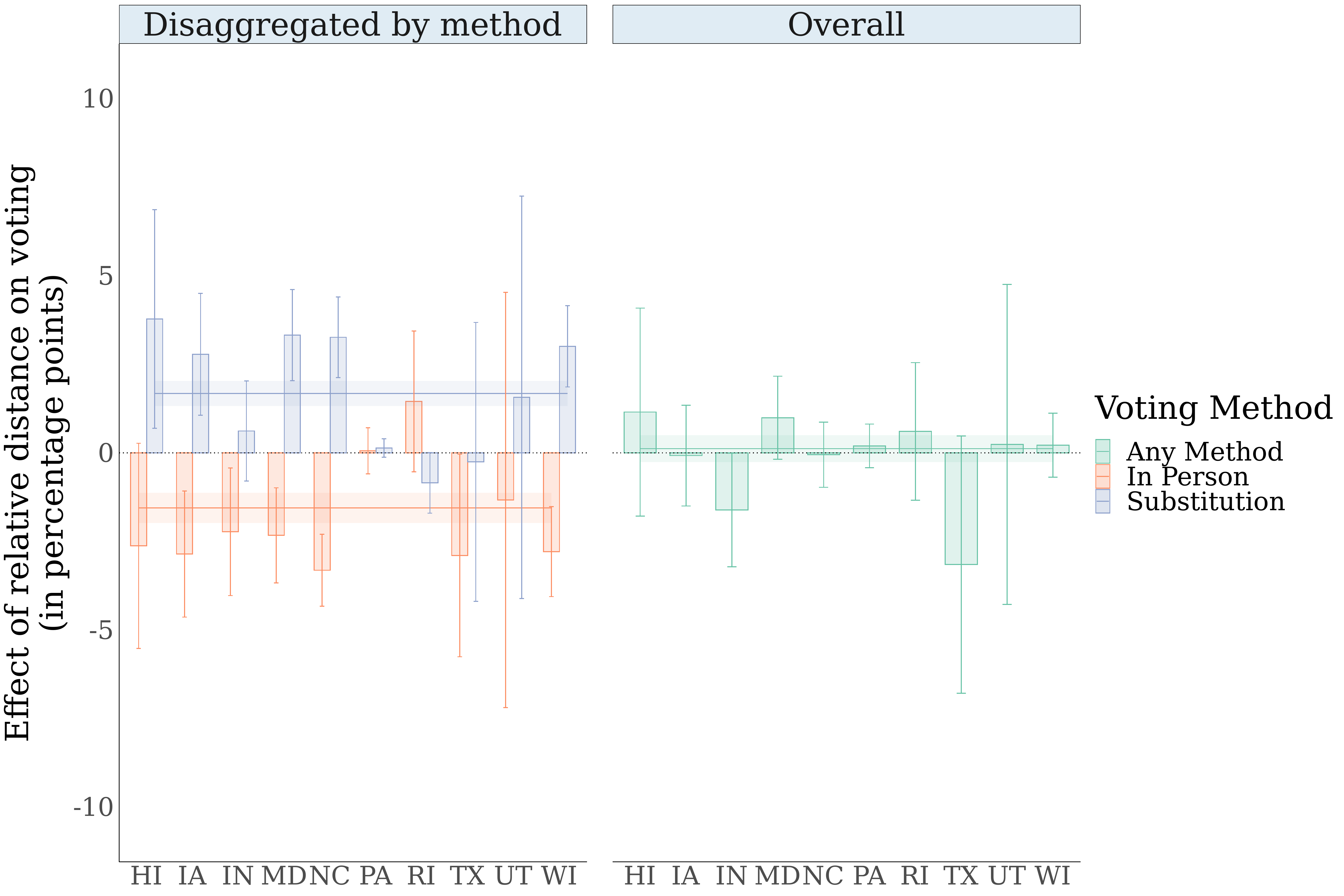}
    \caption{
    Relative distance does not affect the likelihood of voting, but does affect the method used to vote. 
    \label{distance_all}}
\end{figure}

We first report the results of the \distance{}. The left panel of \figref{distance_all} shows that \units{} that live relatively further from their assigned polling places are generally less likely to vote at their assigned polling place on Election Day than \units{} that live relatively closer. The solid orange bars show that in eight of the tens states we study, \units{} who live relatively further from their assigned polling place were less likely to vote at that polling place on Election Day than \units{} who live relatively closer. 
The error bars represent 95 percent confidence intervals for each state as well as our aggregate estimate.
Pooling over all ten states, a \unit{} that lives relatively further from their assigned polling place is 1.6 ($\pm 0.4$) percentage points \footnote{We report 95\% confidence intervals.} less likely to vote at their assigned polling place on Election Day than a \unit{} that lives relatively closer. 

The remainder of \figref{distance_all} shows that \units{} that live relatively further from their assigned polling location compensate for a lower rate of voting at their assigned polling place with higher rates of early in-person voting or mail-in balloting. The solid purple bars show that in eight of the tens states we study, \units{} who live relatively further from their assigned polling place were more likely to use early in-person voting or mail-in balloting than \units{} who live relatively closer. Pooling over all ten states, a \unit{} that lives relatively further from their assigned polling place is 1.7 ($\pm 0.36$) percentage points more likely to use early in-person voting or mail-in balloting than a \unit{} that lives relatively further. The right panel shows that there is almost no difference in the turnout rate of \units{} who live relatively further or closer from their assigned polling place.

\begin{figure}[H]
\begin{center}
    
    \includegraphics[width=.5\columnwidth]{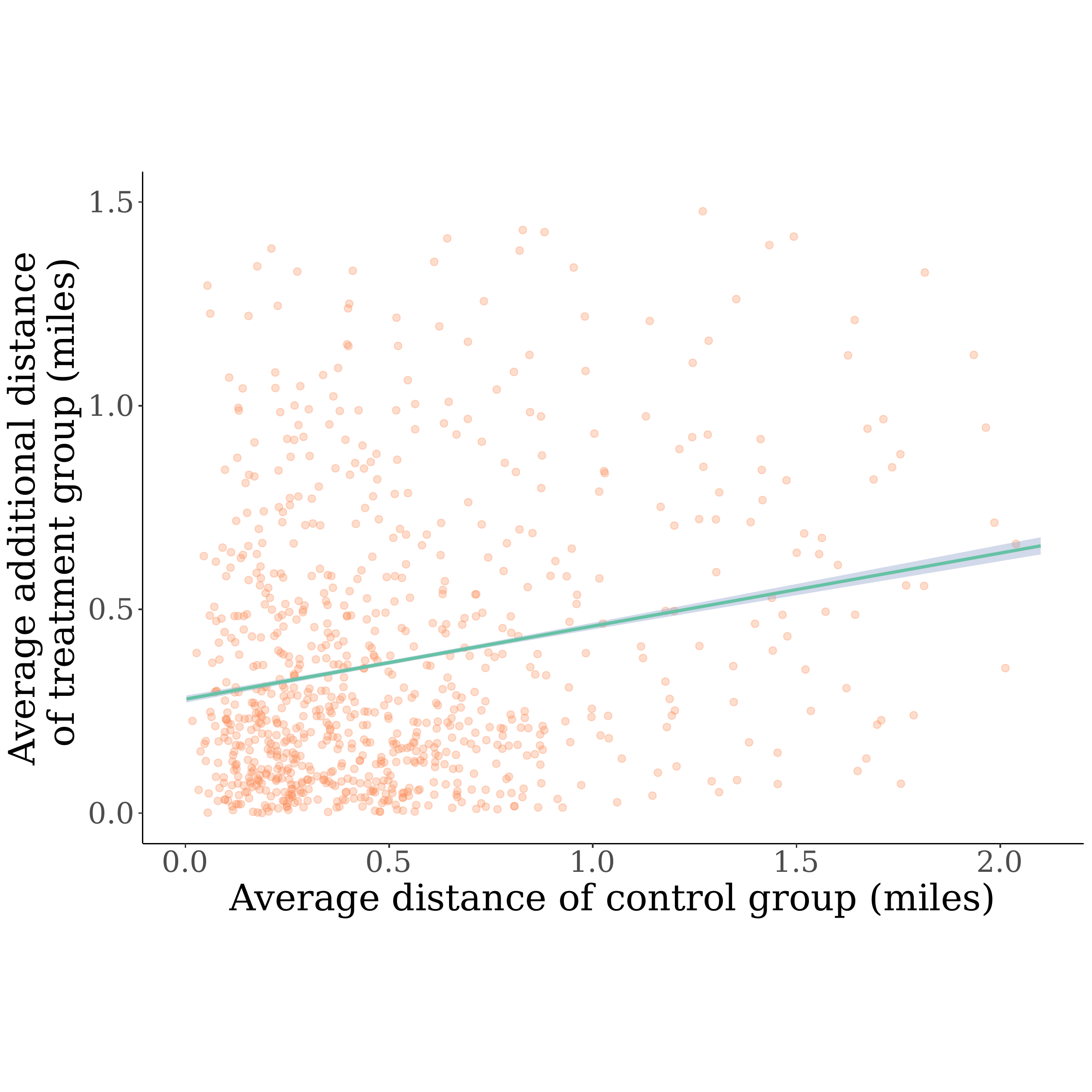}
    \caption{Average distance to the polling place among \units{} who live relatively closer and the average additional distance \units{} who live relatively further must travel in a sample of 1,000 blocks in our \textit{\distance{}} analysis.
    \label{distance_hist}}
    \end{center}
\end{figure}

\figref{distance_hist} helps to contextualize how much further \units{} who live relatively further from their assigned polling location have to travel in order to vote in-person on Election Day. \figref{distance_hist} shows the average additional distance these \units{} must travel as compared to \units{} who live relatively closer in a random sample of 1,000 blocks included in our analysis. Even though 85\% of all distance differences are less than one mile, this is still sufficient to generate the effect of relative distance on both voting \ip{} and by \sub{} observed in \figref{distance_all}. \figref{distance_hist} also demonstrates that the additional distance that \units{} that live relatively further from their assigned polling location must travel to that polling location increases as a function of the distance that \units{} that live relatively closer to their assigned polling location must travel to that polling location. 

\begin{figure}[H]\centering
    \includegraphics[width=.55\columnwidth]{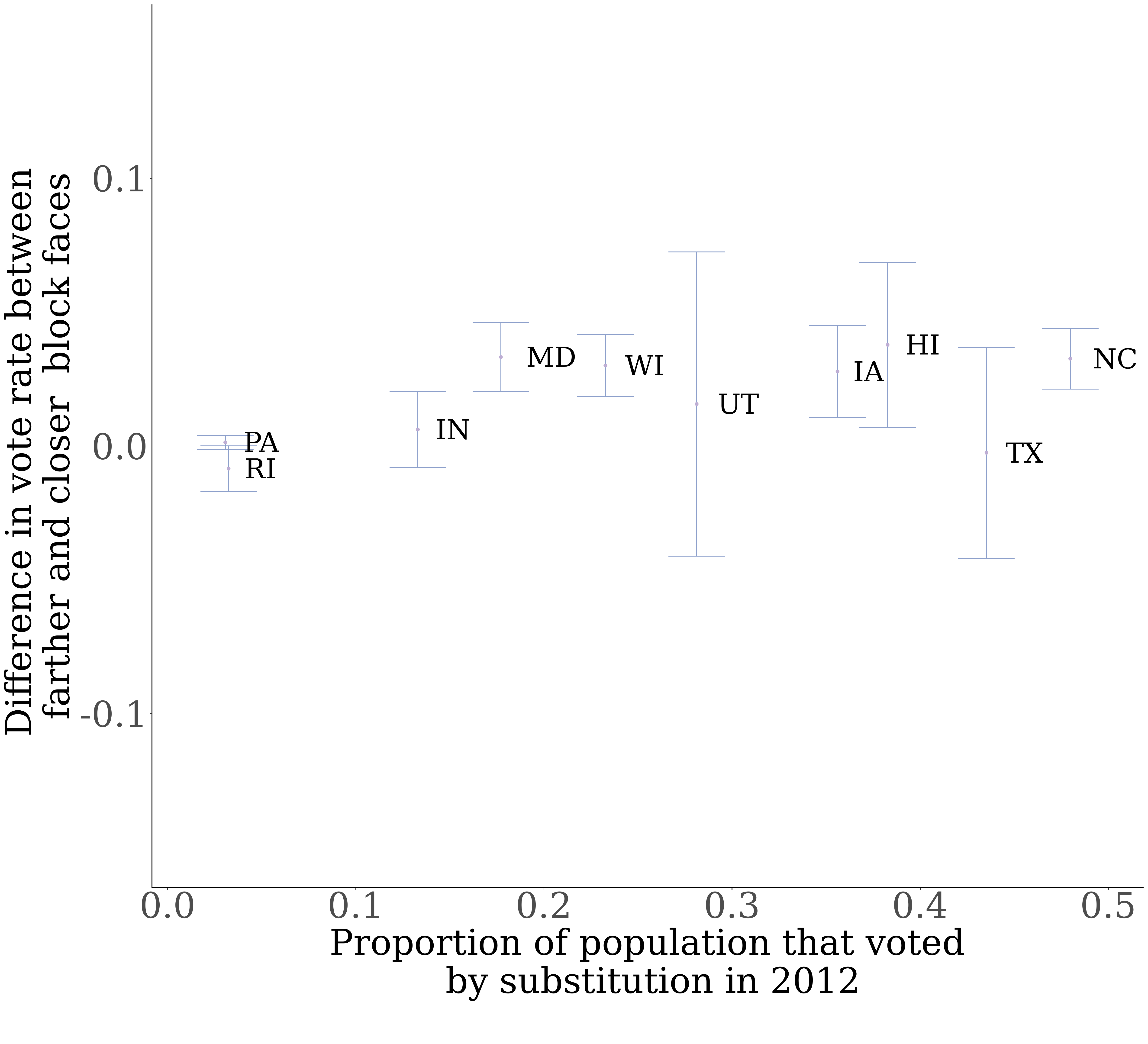}
    \caption{Registrants who live relatively further from their assigned polling location are least likely to respond with substitution in states with the lowest usage of mail or early in-person ballots. 
    \label{eavs}}
\end{figure}

\figref{distance_all} shows that substitution methods appear to offer an alternative to \units{} when the cost of voting on Election Day increases. However, access to substitutions varies between states and even within states. For example, some states allow for both no-excuse mail ballots and early in-person voting  (e.g., North Carolina), while others allow no-excuse absentee voting only for certain voters (e.g., Indiana). Thus we inspect how voting by substitution varies with a state's openness to substitution adoption. As a proxy for a state's openness to substitution adoption, we use the percentage of voters that cast a mail or early in-person ballot in the 2012 presidential election.
Consistent with our expectations, \figref{eavs} shows that \units{} who live relatively further from their assigned polling location are least likely to respond with substitution in states with the lowest level of adoption. However, we cannot rule out the possibility that there may be other state-specific factors that affect substitution patterns and the use of mail or early in-person ballots.

\begin{figure}[H]\centering
    \includegraphics[ width=.65\columnwidth]{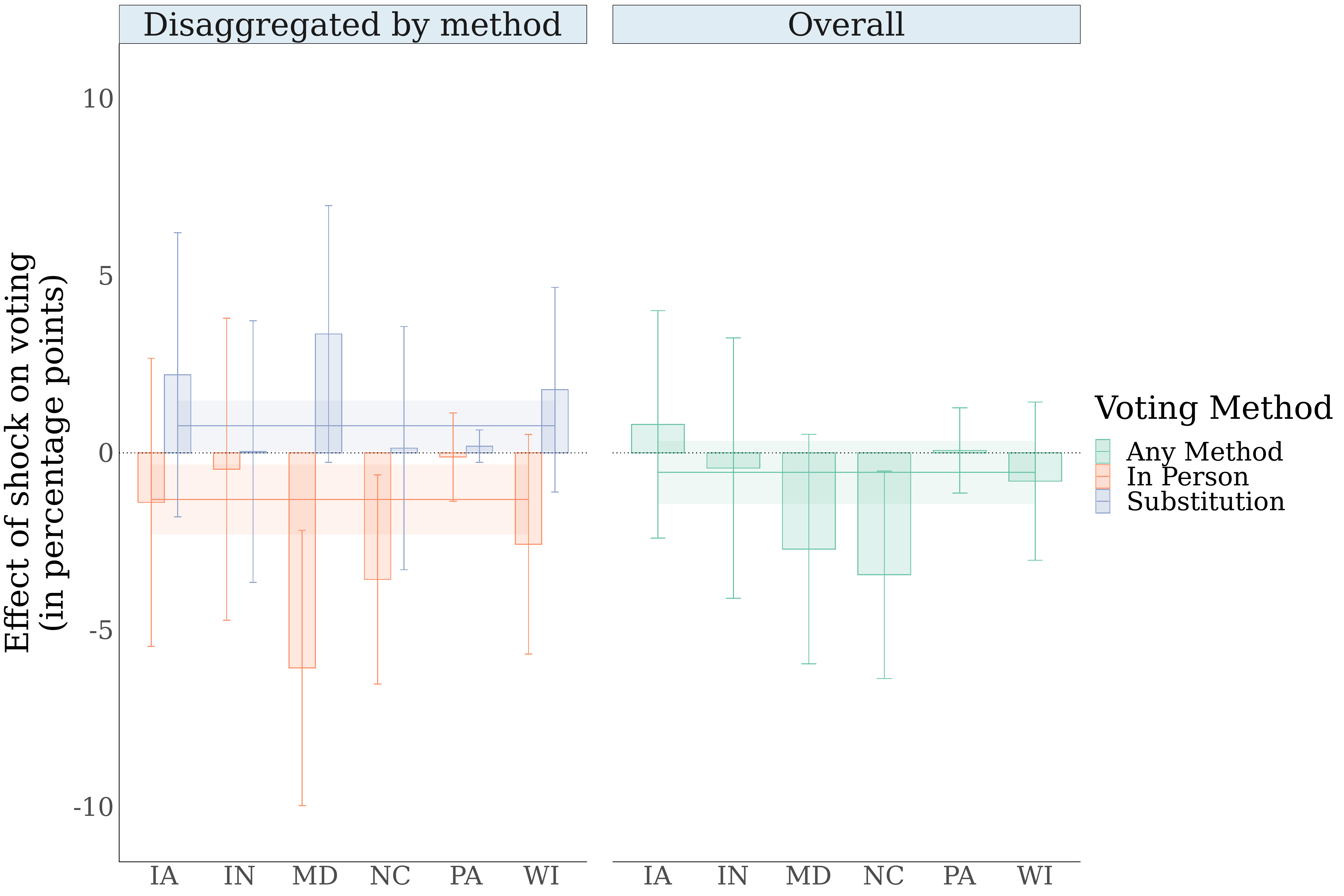}
    \caption{
    While shocks reduce Election Day voting, more that half of this reduction is compensated for by increased voting by substitution. 
    \label{shock_all}}
\end{figure}

Next, we address the question of how a shock affects Election Day voting, voting by substitution and voting by any method. In \figref{shock_all} we consider the data produced by the \shock, where one block face experiences a polling place assignment shock in 2016. It shows that shock creates a shift in how voting occurs, reducing voting in person by 1.3 ($\pm 1.00$) percentage points and increasing voting by substitution by 0.80 ($\pm 0.73$) percentage points. 
Overall, we see that as with \distance{}, \shock{} is moderated by \units{} voting by substitute modalities rather than in person.

\begin{figure}[H]
\begin{center}
    
    \includegraphics[width=.5\columnwidth]{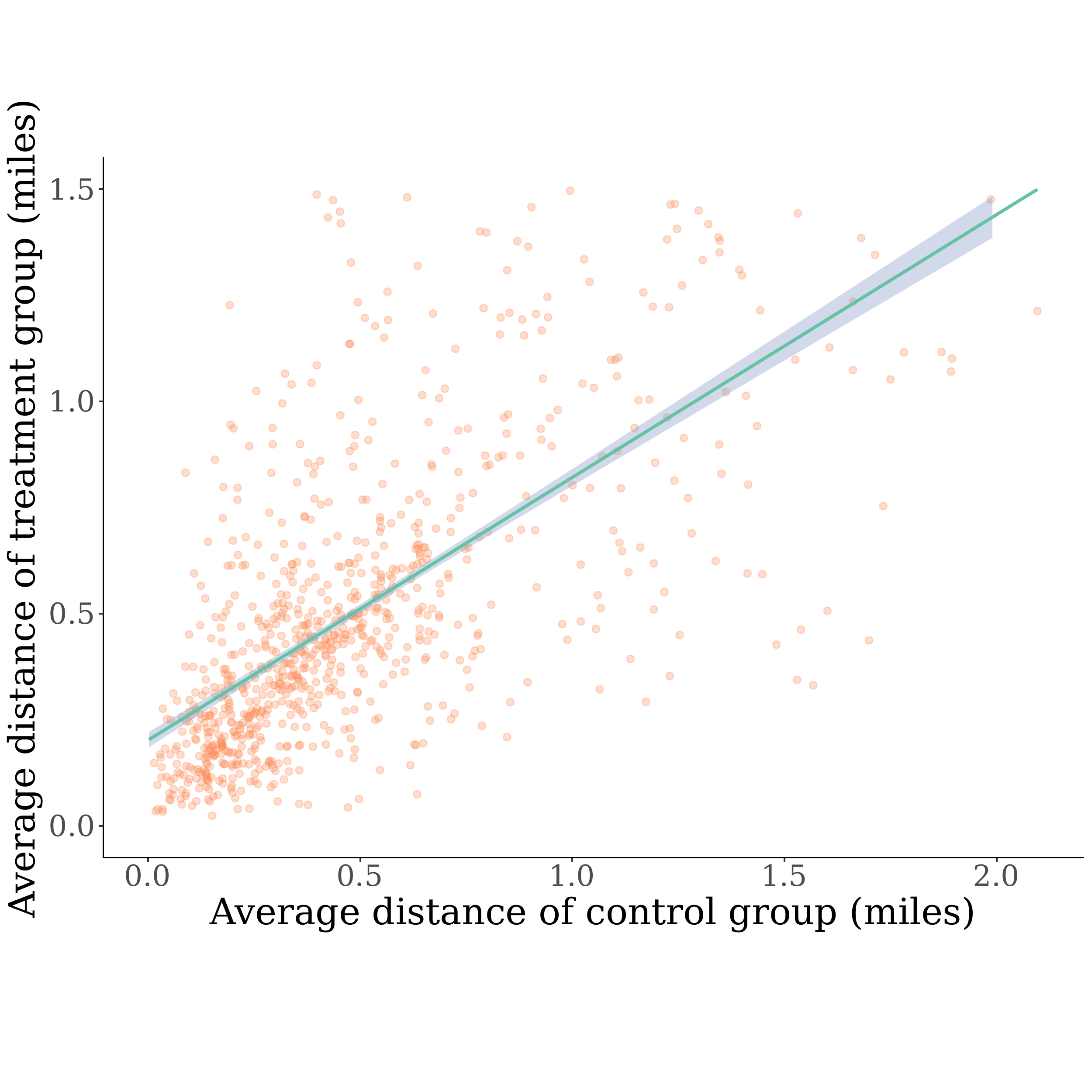}
    \caption{Average distance to the new polling place among \units{} who experience a polling place change in a sample of 1,000 blocks in our \textit{\shock{}} analysis.
    \label{shock_hist}}
    \end{center}
\end{figure}

One complication with examining shocks is that \units{} experiencing shocks often also  experience changes in the relative distance between their residence and their polling place. \figref{shock_hist} shows that some shocks cause \units{} to reside closer to their polling place, while other shocks cause \units{} to reside further. One reason for this heterogeneity is that the shocks in our data result from a combination of counties adding polling places, consolidating polling places, and moving polling places. Each of these events might influence the cost of voting differently. Consolidations may not only increase the average distance \units{} travel to their polling place, but also increase the cost of voting via longer lines. Conversely, additions may make Election Day voting more convenient in other ways beyond reducing the average distance \units{} travel to their polling place. \tabref{types_of_shocks} in the Appendix shows that more \units{} who experience a shock live in a county that added polling places than a county that subtracted polling places, suggesting that additions may be more likely to generate the shocks in our data than subtractions. 

\begin{figure}[H]\centering
    \includegraphics[ width=.95\columnwidth]{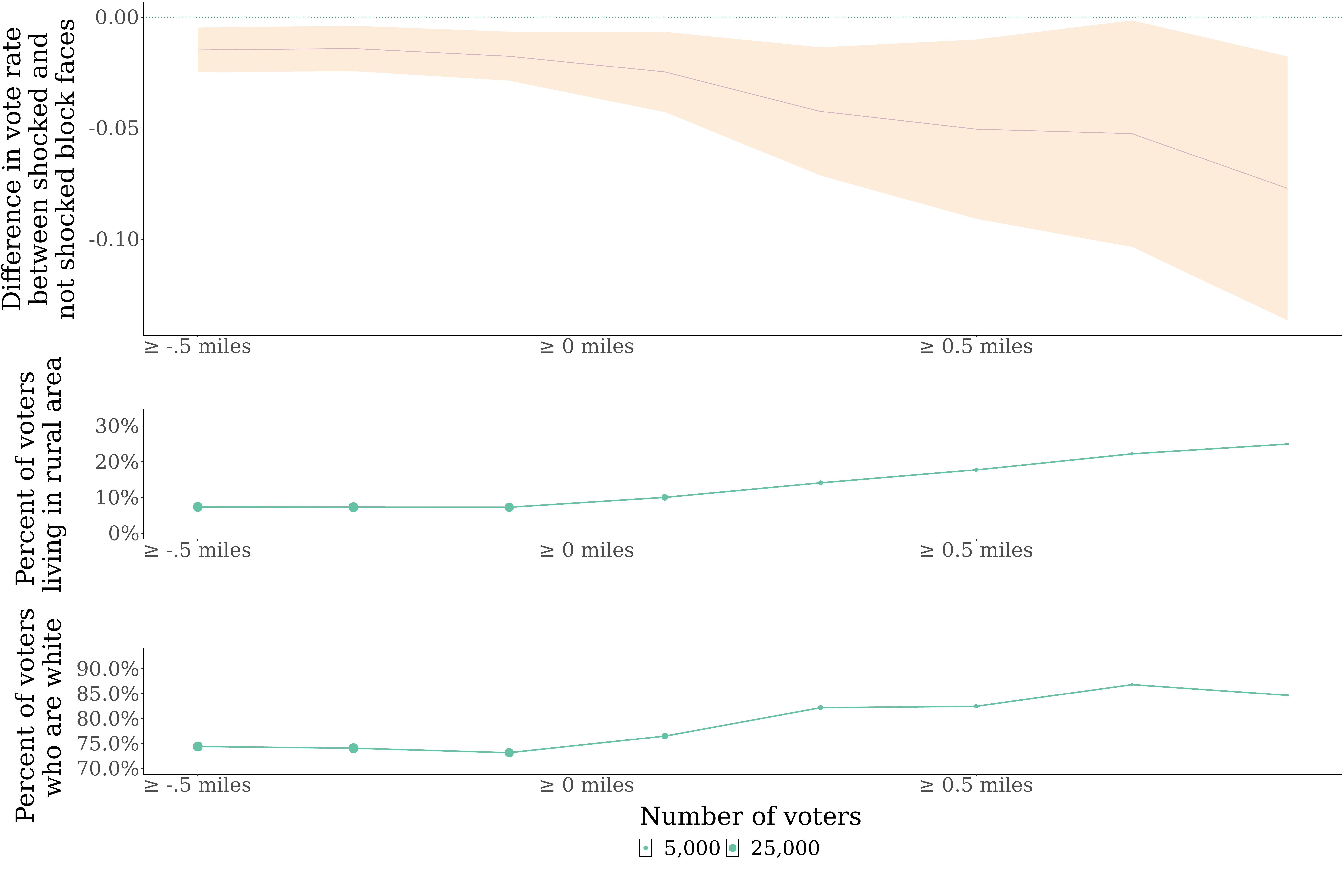}
    \caption{
    Shocks cause a greater reduction in Election Day voting when the sample of blocks is restricted to those where difference in distance to the polling place is greater among those who did and did not experience a shock. 
    \label{shock_windows}}
\end{figure}

\figref{shock_windows} examines whether the \shock{} on Election Day voting is more pronounced when \units{} must travel greater distances than they were previously required. It inspects the difference in the voting rate between the block side which experienced a shock and the side that did not as a function of the difference in the distance of the block from the new polling place and the distance of the block from the old polling place. The \(x\)-axis varies which blocks are included when estimating the \shock{} based on this difference in distance. For example, the estimate reported for ``$>$ 0.5 miles'' only includes blocks that experienced a shock which resulted in the new polling place being 0.5 miles or more further from the block than the old polling place. The top panel shows that the relatively further the new polling place is from the block, the greater the decline in Election Day voting among those that experience the shock. Similarly, Figure \ref{diff_distances} in the Appendix shows that the \distance{} on Election Day voting is larger the more we restrict the sample to blocks where the difference in distance between the further and closer polling place is larger. 

The trend demonstrated in \figref{shock_windows} cannot necessarily be interpreted as the effect of distance. The middle and bottom panels show that the underlying population used to construct each estimate varies, with shocks causing larger differences in distance to the polling place in shocked blocks in rural areas. Moreover, shocks that cause \units{} to reside substantially further from their polling place are disproportionately likely to be caused by consolidations.
We leave it to future work to  isolate the simultaneous effect of a unit increase in distance on \units{} experiencing polling place assignment shocks.

\section{Discussion}
While existing designs rely on knowledge of precinct boundaries, ours instead utilizes residential addresses and polling place addresses. 
This
novel quasi-experimental design allows us to utilize data from multiple states and present the most comprehensive analysis to date of the \distance{} and \shock{}.   We find clear evidence that both cause \units{} to cast early in-person or mail-in ballots instead of voting at their polling place on Election Day. 
The reduction in Election Day voting caused by having a longer trip to the polls appears to be completely offset by increases in early in-person or by mail-in voting. Likewise, our point estimates suggest that a majority of the reduction in Election Day voting caused by shocks are offset by increases in early in-person or by mail-in voting, and that overall \units{} who experience shocks are similarly likely to vote as those who live on the same block who do not.

The finding of minimal effects of shocks on overall turnout contrasts with many of the previous studies of shocks. We speculate that one reason why our findings differ from previous work is that most previous work focuses specifically on shocks resulting from the consolidation of polling places. In contrast, we find that in our data \units{} are more likely to experience shocks in counties that are adding polling places than in counties that are reducing polling places, suggesting that many of the shocks in our data are not caused by consolidation. While all shocks impose search costs, shocks that result from consolidation are likely to also impose other additional costs on voting. Conversely, shocks that result from polling place exchanges, or especially additions, may reduce the cost of voting in other ways. 

Our results also contrast with \citet{cantoni2020precinct}, which is the only other paper that studies the \distance{} independent of shocks. While \citeauthor{cantoni2020precinct} finds a substantial decrease in overall turnout among \units{} located further from their polling place relative to similarly situated \units{} that are located closer, we find \units{} who need to travel further to their polling place compensate for their reduced Election Day voting with greater early in-person and mail voting. There are a number of potential explanations for why we reach different conclusions. First, the ability for \units{} to switch to early in-person or mail balloting likely depends, in part, on the accessibility of these alternative voting methods. Second, \units{} may not respond differently to small and large differences in relative distance. Most blocks in our sample were within a mile of their assigned polling place and the \units{} who lived further from their polling place usually lived less than 0.5 miles further from their polling place than the \units{} who lived closer. Finally, the \distance{} may depend on the characteristics of the specifics \units{} and elections being studied. \citeauthor{cantoni2020precinct} finds, for example, the \distance{} is greater in areas where car ownership is lower, and is lower in the 2016 presidential election than in the other elections that he studied. 

While our identification strategy can be applied broadly in settings where \units{} are assigned to polling places, it is limited in some respects. First, it only allows us to estimate the \distance{} and the \shock{} on the turnout of \units{}. If the experience of greater distance or shocks causes potential voters to not be registered, then our design may underestimate the effect of these variables on turnout \citep{NyhanSkovronTitiunik2017}. Moreover, our focus on \units{} who lived at the same address for four years means our sample under-represents the \distance{} and the \shock{} on potential voters with less residential stability. We also require that a \unit{}'s address can be reliably geocoded. Perhaps because rural addresses might be more difficult to reliably geocode and we require all \units{} on a block to live within 0.3 miles of each other, \units{} from rural areas appear to be underrepresented in our analysis. 

Future work can build off the identification strategy we propose here to develop more complex measures of distance. Here, we simplify all measures of continuous distance into a simple notion of farther or closer. This distinction allows us to compare \units{} across geographies in a way which we believe is statistically defensible. However, this means our estimated treatment effects bundle together many comparisons of the \distance{} that likely affect voting behavior in different ways. However, we do not think it necessarily makes sense to bundle together blocks with similar values of distance differences because a given distance difference may correspond to a large increase in the cost of voting in a urban areas where people walk to a polling place and a small increase in the cost of voting in a rural area where everyone drives to the polls. We believe future work would benefit from greater thinking about how to translate measures of relative distance in different contexts into a measure of the time and effort that is required to reach a polling place.

Our work is motivated by a desire to understand the magnitude of the costs of voting on Election Day. While we caution that these are limited to the population of registered voters, an optimistic result of our work is that these costs are readily offset by the ability of \units{} to shift into substitute modalities. The provision of mail-in ballots and early     voting venues can have important implications beyond the ability to vote, but perhaps on the experience of political engagement. We leave it to future work to explore 
 not only the ability of
various voting modalities to offset costs but the implications of this effect for all potential voters.

\bibliographystyle{plainnat}
\bibliography{references}

\appendix
\renewcommand\thefigure{\thesection.\arabic{figure}}    
\setcounter{figure}{0}  
\renewcommand\thetable{\thesection.\arabic{table}}    
\setcounter{table}{0} 
\clearpage
\section{Appendix}
\subsection{Additional Figures}
\label{sec:additional}
\begin{figure}[H]

\centering
\resizebox{.9\linewidth}{!}{
    \includegraphics[ width=.99\columnwidth]{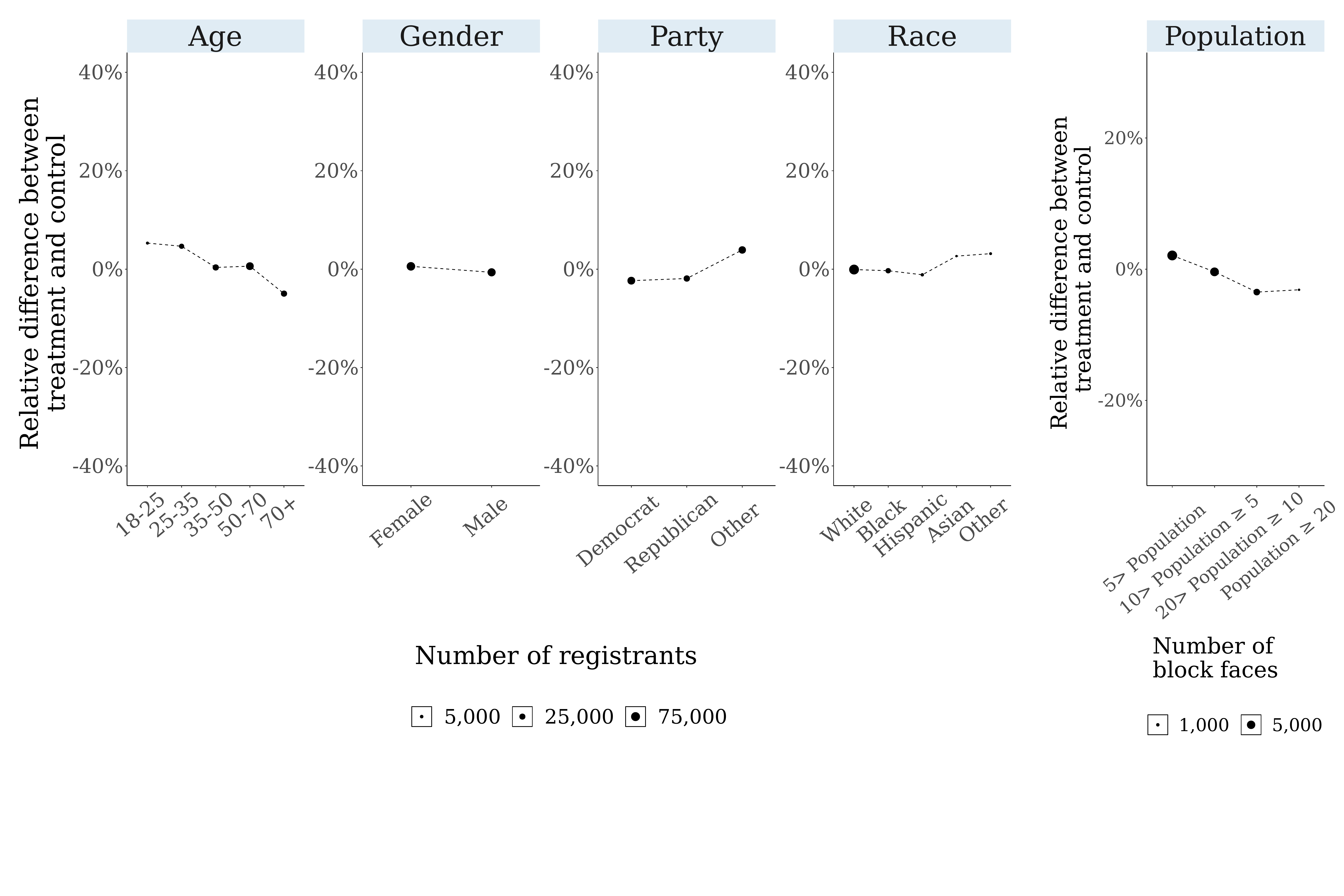}}
    \caption{We see the age, gender, modeled partisanship, modeled race of \units{}, as well as indicators representing that the total number of \units{} are similar on either block face. (Distance)
    } 
    \label{bp_distance}
    
\end{figure}

\figref{bp_distance} shows how the demographics of registrants who live on the side of the block that is further from its polling place (i.e., treatment group) compares to the demographics of registrants who live on the side of the block that is closer to its polling place (i.e., control group). We inspect age, gender, modeled partisanship, modeled race, as well as the number of \units{}
on either side of the block. 
 Each point is the difference between the proportion of registrants with a given characteristic in treatment and the proportion of registrants with this characteristic in control, divided by the proportion of registrants with a given characteristic in control. The size of each point is scaled by the number of registrants in the control group, where larger points represent groups with more registrants. 
 
 The difference between the proportion of registrants with a given characteristic in the treatment group relative to the control group is close to zero for the most common characteristics. For less common characteristics the differences can be greater. Here, we show the percent differences. However, note that all absolute differences are less than 2 percentage points and that the majority are less than one percentage point.

\begin{figure}[H]\centering
\resizebox{.45\linewidth}{!}{
    \includegraphics[width=.99\columnwidth]{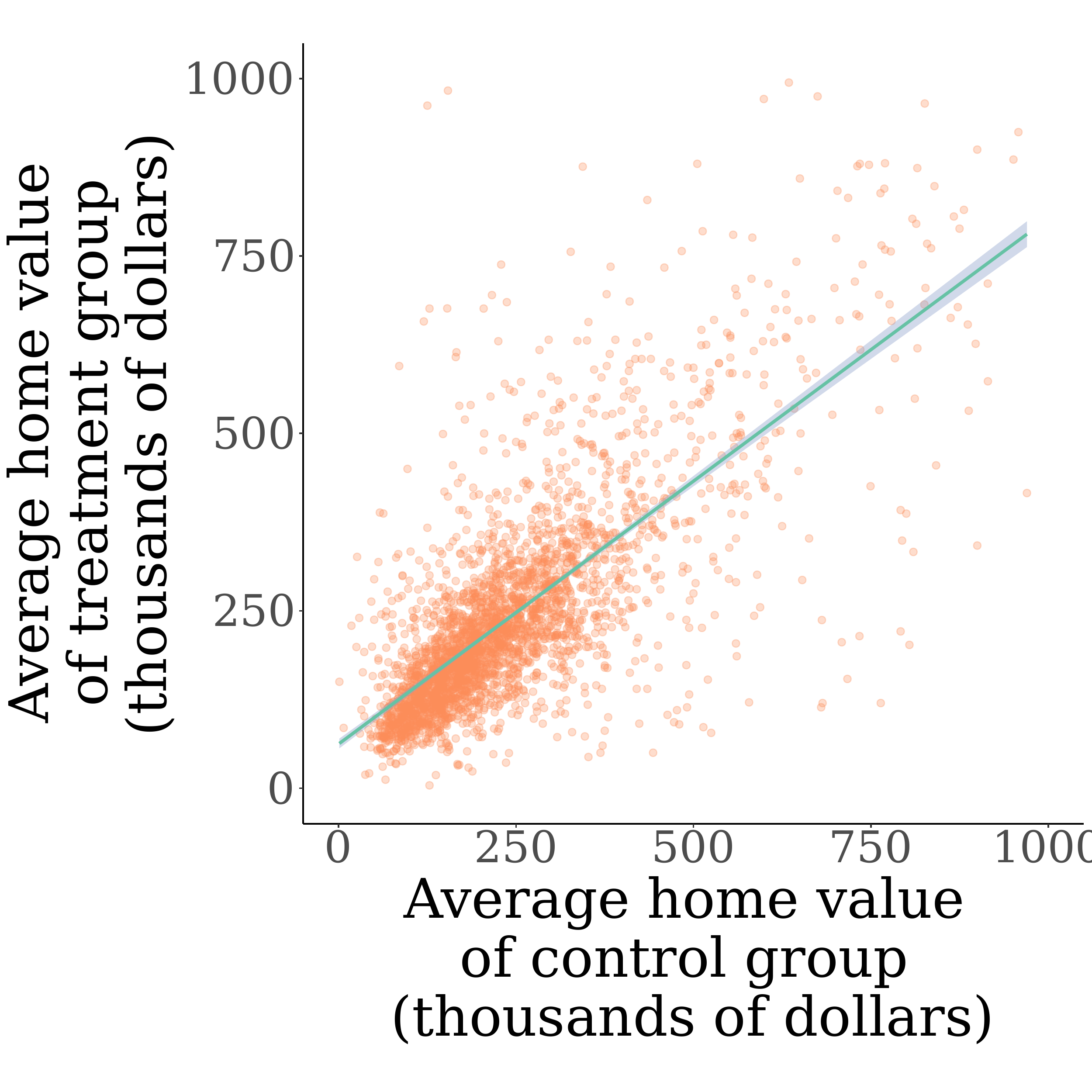}}
    \caption{
   We see that the home prices of \units{} on either block face are similar. (Distance). 
    \label{scatter_real_estate_distance}}
\end{figure}

\begin{figure}[H]
\centering
\resizebox{.43\linewidth}{!}{
    \includegraphics[width=.99\columnwidth]{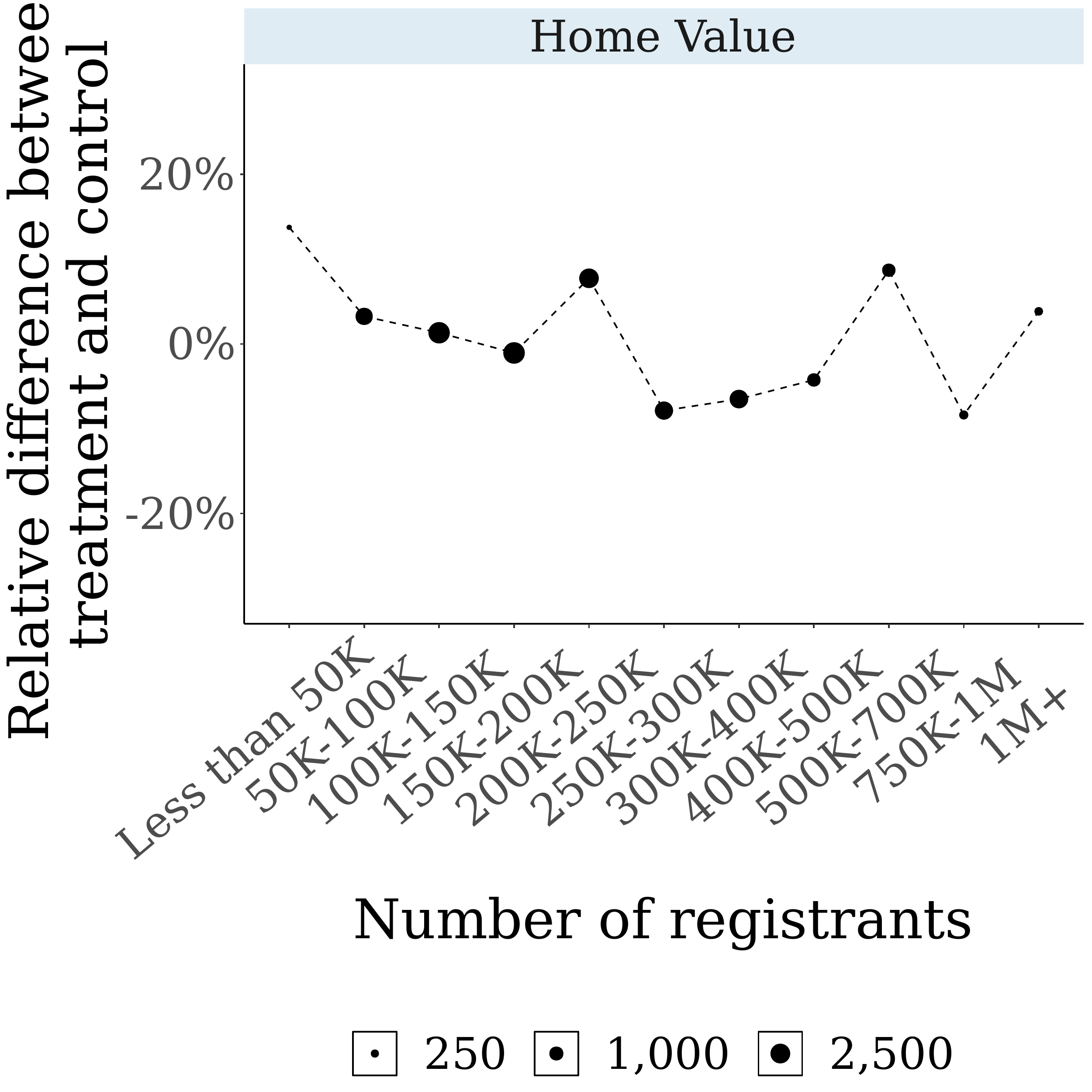}}
    \caption{
   We see that the home prices of \units{} on either block face are similar. (Distance)
    \label{balance_real_estate_distance}}
\end{figure}

We also observe information on the sale prices of homes for about 14 percent of registrants.  \figref{scatter_real_estate_distance} and \figref{balance_real_estate_distance} show that houses sell for a similar amount in the treatment and control groups for this subset of \units{}. Home value amount is estimated based on a number of public record data, such as documents filed at the county recorder's office.

\begin{figure}[H]\centering
\resizebox{.9\linewidth}{!}{
    \includegraphics[ width=.99\columnwidth]{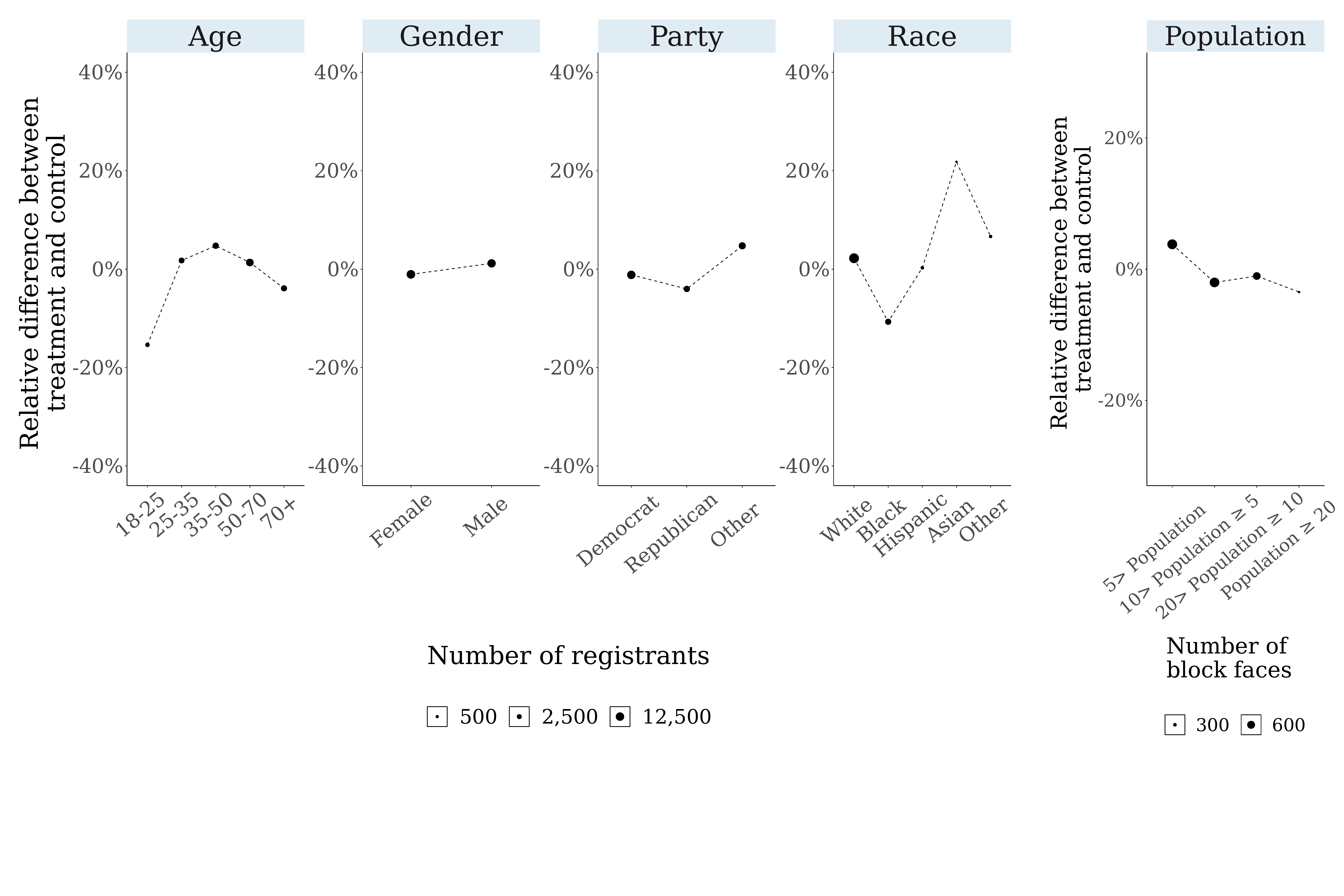}}
    \caption{We see the age, gender, modeled partisanship, modeled race of \units{}, as well as indicators representing that the total number of \units{}, are similar on either block face, although the percent difference can be larger because the sample size is smaller. (Shocks)
    } 
    \label{bp_shock}
\end{figure}

\begin{figure}[H]\centering
\resizebox{.4\linewidth}{!}{
    \includegraphics[width=.99\columnwidth]{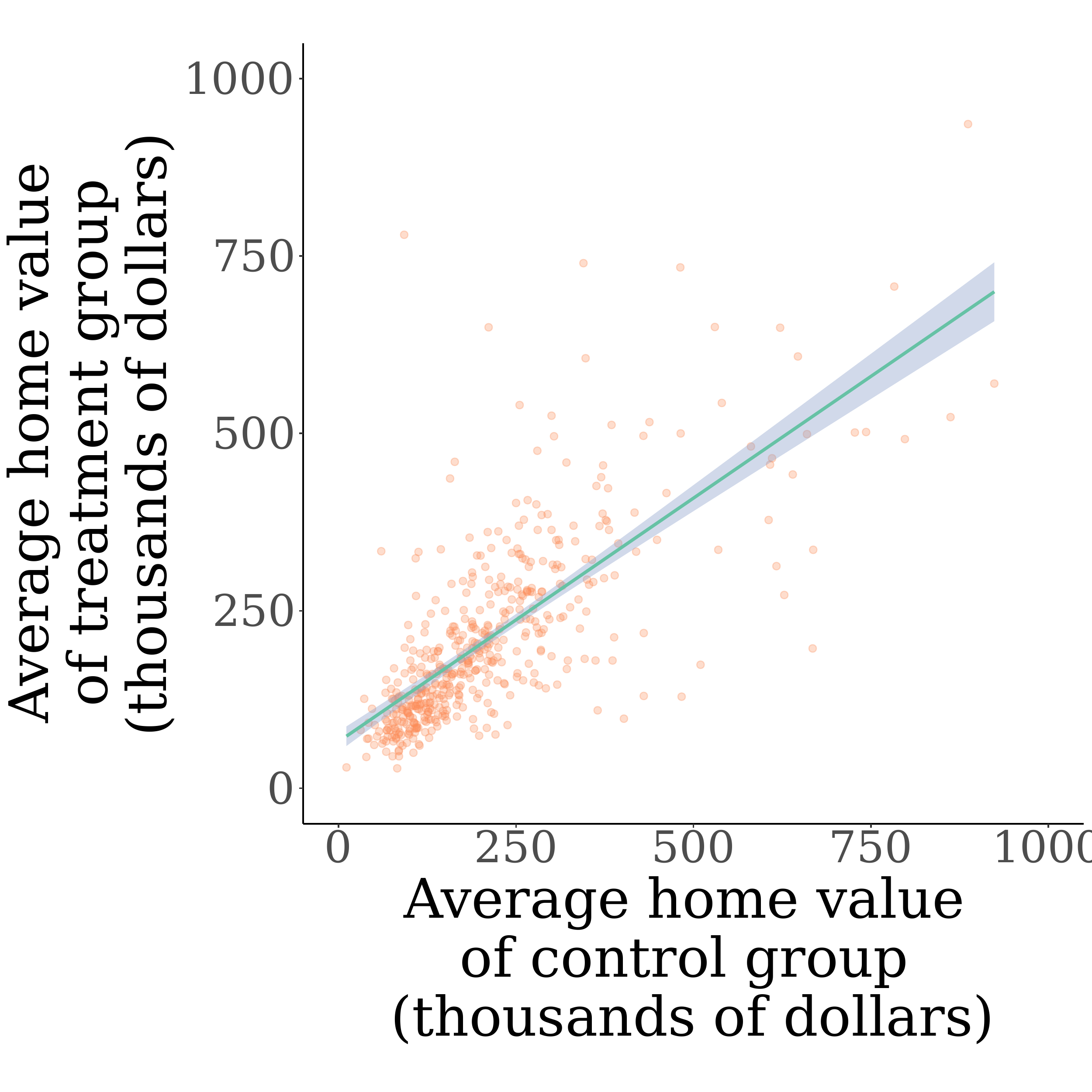}}
    \caption{
   We see that the home prices of \units{} on either block face are similar. (Shock)
    \label{scatter_real_estate_shock}}
\end{figure}

\begin{figure}[H]\centering
\resizebox{.43\linewidth}{!}{
    \includegraphics[width=.99\columnwidth]{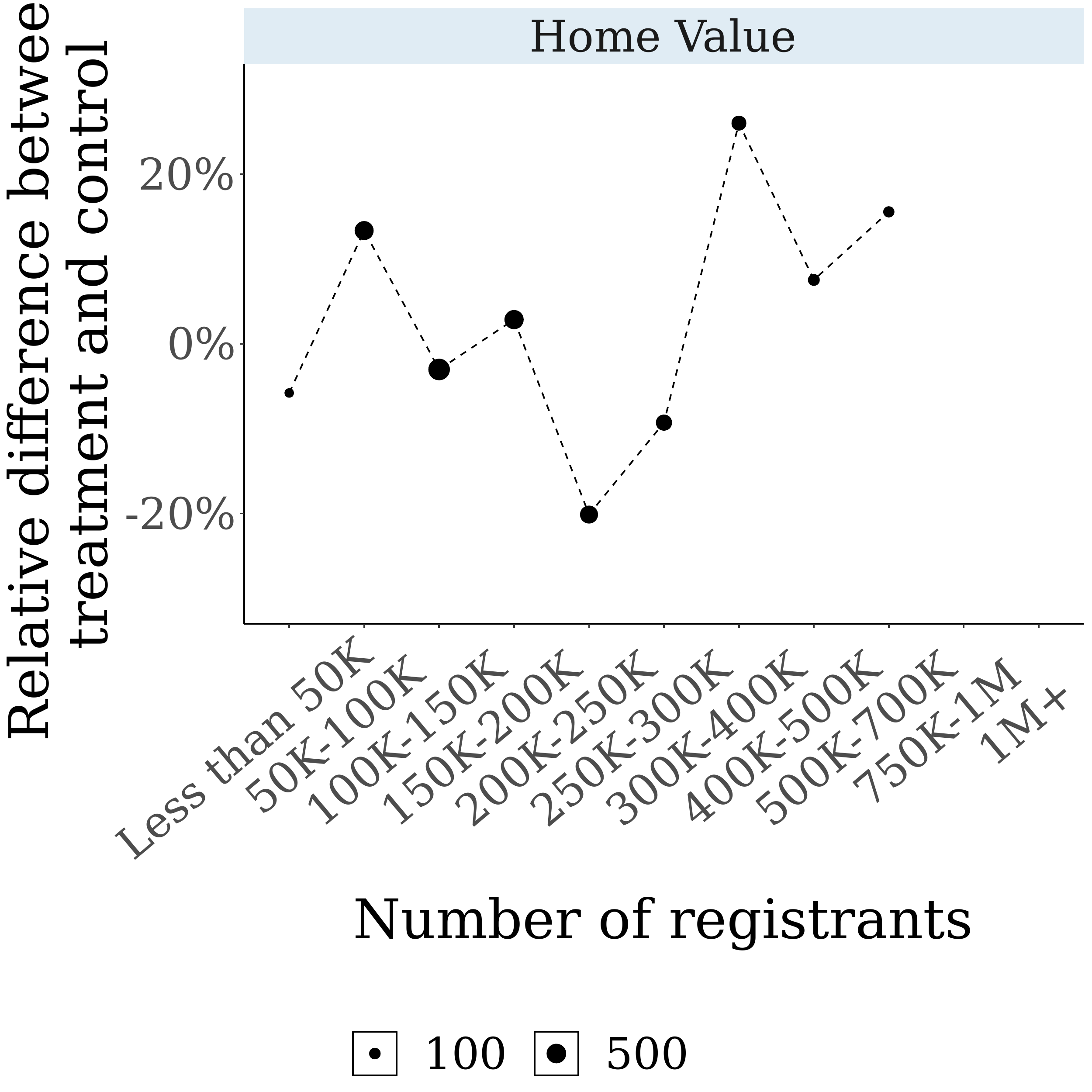}}
    \caption{
   We see that the home prices of \units{} on either block face are similar. (Shock)
    \label{balance_real_estate_shock}}
\end{figure}

\begin{figure}[H]\centering
\resizebox{.45\linewidth}{!}{
    \includegraphics[width=.99\columnwidth]{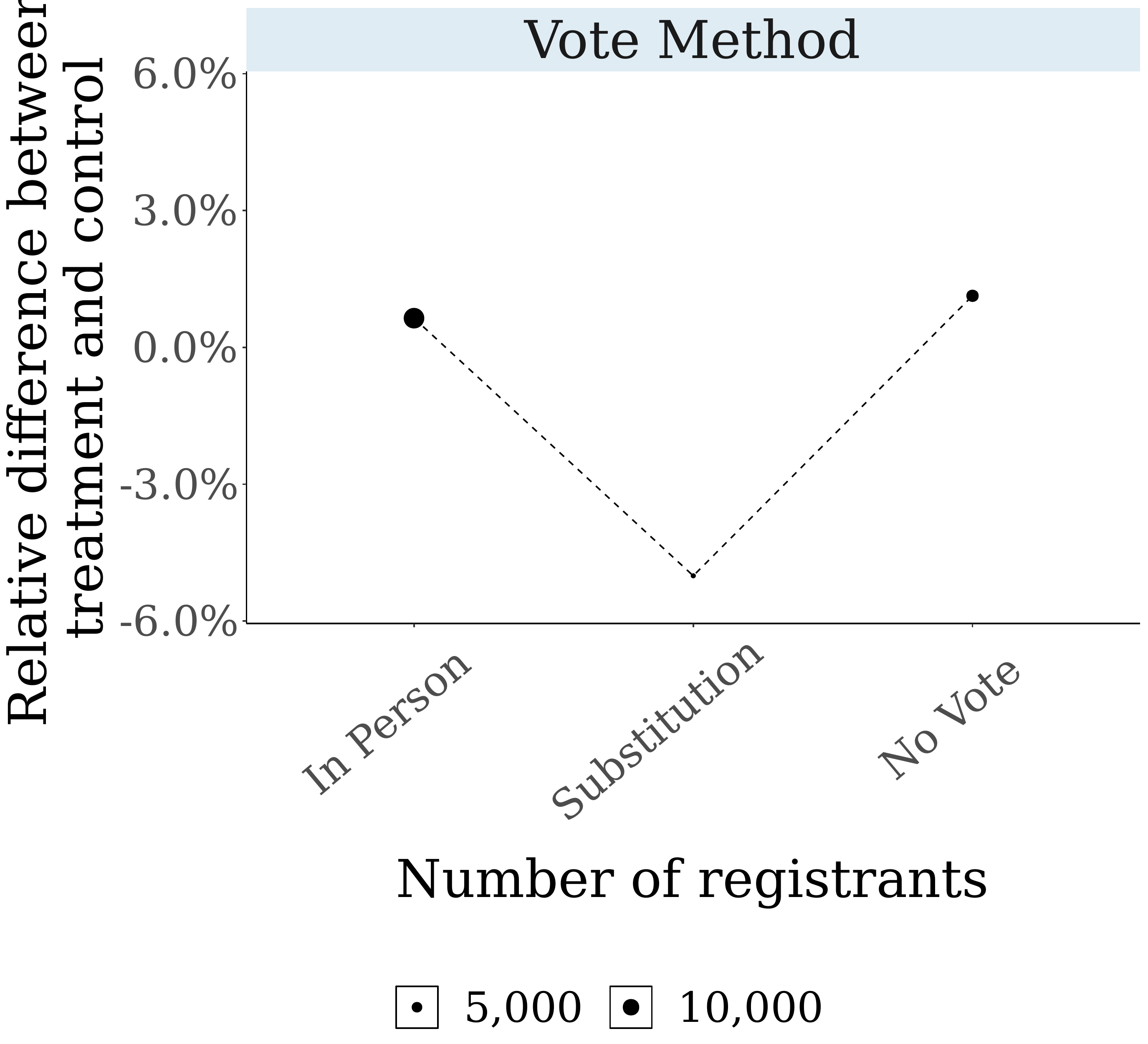}}
    \caption{
   We see that the historical voting patterns in the 2012 presidential election of \units{} on either block face are similar, with \units{} experiencing a shock being, if anything, less likely to vote by substitution than \units{} who did not. The roughly 5 percent lower usage of substitution voting reflects that 13.1 percent of \units{} who experienced a shock voted using a substitution method in 2012 as compared to 13.8 percent of \units{} who did not.
    \label{balance_past}}
\end{figure}

Next, we conduct the same analysis except defining those on the side of a block experiencing a polling place change as the treatment group and those of the side of a block keeping the same polling place as the control group. The one notable difference is that, because we have substantially fewer observations, \figref{bp_shock} sometimes reveals larger percent differences than were observed in \figref{bp_distance}. Once again  \figref{scatter_real_estate_shock} and \figref{balance_real_estate_shock} show housing prices are comparable on the side of blocks that do and do not experience shocks for the 12 percent of registration records for which that information is available. Unlike with the relative distance, we also expect to observe similar voting patterns in the treatment and control groups in the previous election before the shock was realized, which \figref{balance_past} shows is the case.

\begin{figure}[t]
\begin{center}
    \includegraphics[width=.85\columnwidth]{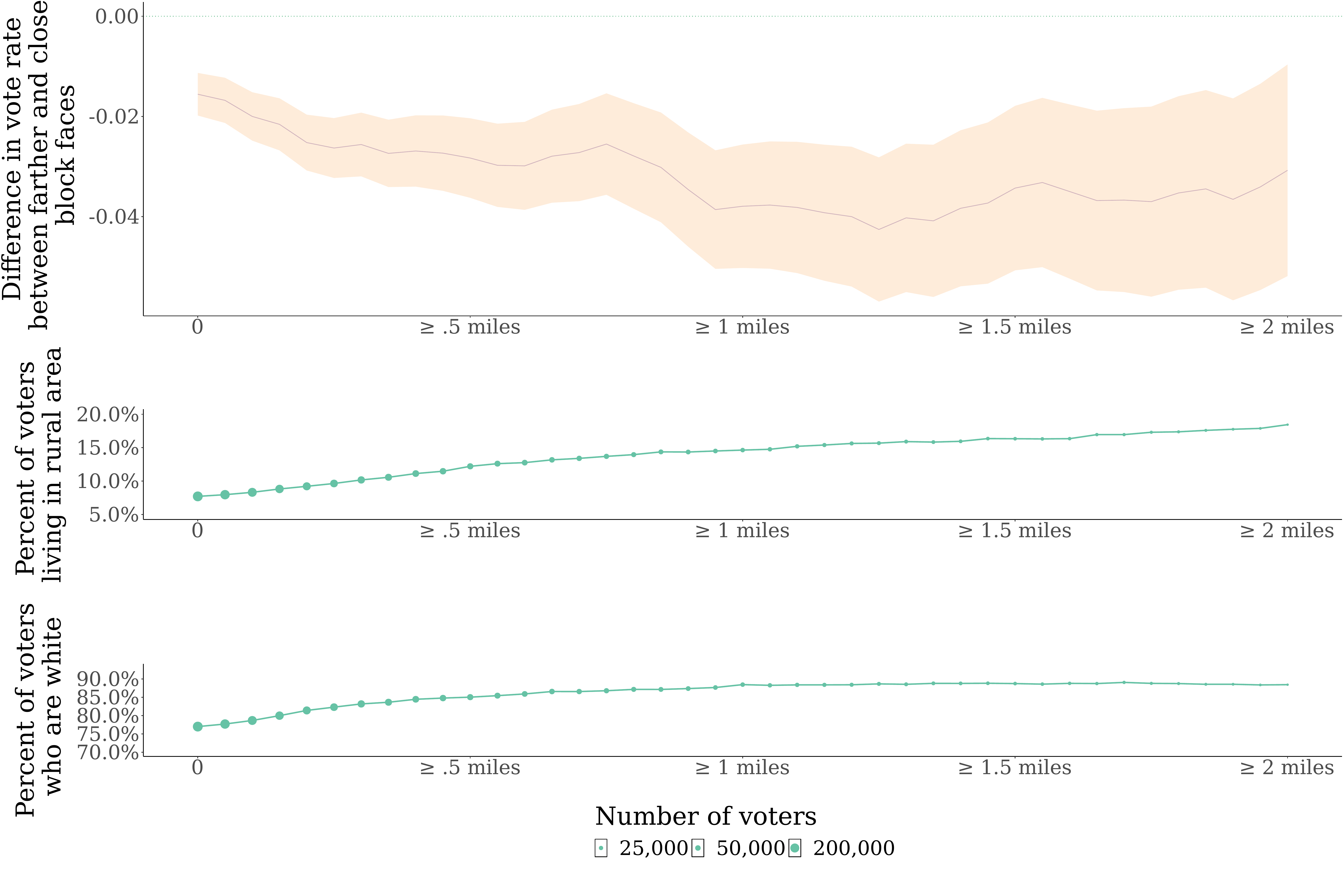}
    \caption{As the difference in distance between the closer and farther block faces grows, so does the effect of distance. However, the population also changes with distance.\label{diff_distances}  }  
   \end{center}
\end{figure}

In \figref{diff_distances} we show the effect of relative distance for all blocks where the additional distance traveled by the treatment group ranges from 0 to more than 2 miles. When we include all voters we see that the average distance is quite small and the effect of relative distance is around 2 percentage points. While we see a more pronounced effected of relative distance for those voters who live on a block with a relatively large difference between the distances traveled by the close and far block faces, this might be attributable to other characteristics of these voters which differentiate them from the rest. For example, they tend to be more rural and more white.  

\subsection{Types of shocks}
\label{sec:shock_extra}
Polling-place assignment changes can occur
 when precinct boundaries are redrawn or to meet the changing constraints placed on local election officials, e.g. when new regulations render a former site unfit. 
 Broadly, polling places are 1) removed and consolidated within jurisdiction, 2) added to a jurisdiction and 3) moved to different locations where the total number within a jurisdiction is unchanged. 
To examine the types of shocks experienced by voters in this analysis,  
we  describe each county in the dataset as either removing, adding, or not changing the total number of polling places from 2012 to 2016. \figref{types_of_shocks} shows both the percentage of counties and the percentage of voters living in those counties for each type of change to the number of polling places at the county level. 
We find that in our data the majority of voters who experience a shock live in a county where polling places were moved.
To inspect county level changes, we consider a polling place to be a precinct name, assuming that within a county precinct names are not duplicated. This definition however may be likely to over-estimate the number of polling places as several precincts are often assigned to the same polling place. 

\begin{table}[H]
\resizebox{\textwidth}{!}{%
\centering
\begin{tabular}{|l|r|r|}
\hline
&\multicolumn{1}{|c|}{Percent of counties}& \multicolumn{1}{|c|}{Percent of registrants} \\
\hline
\makecell{Polling places removed in 2016}& 14.7 & 21.5 \\
\hline
\makecell{Polling places added in 2016 } &  41.1& 55.3 \\
\hline
\makecell{Same number of polling places  in 2016 and 2012} & 44.2 & 23.2  \\
\hline
\end{tabular}
}
\caption{ Different types of shocks and the percentage of counts and voters affected. 
\label{types_of_shocks}
}
\end{table}%

 Each of these events might influence voting differently. For example, consolidations leave fewer polling places for voters, increasing the distance for some voters to reach the polling place. They might also increase the cost of voting via longer lines and less staff members at the polling place. 
 Alternatively, new polling places might reduce some costs, such as distance and time at the polling place.
Analysing  these distinct events would likely uncover different effects for each. 
 
 Here, we describe shock broadly, as having one's polling place assignment changed. 
 While this description does not differentiate between the nuanced experiences of each type of change, it captures the common event of experiencing a search cost, where a voter must now determine where to vote. We see that this cost reduces in-person and increases substitution voting, respectively, even when the type of shock is not considered and when roughly 55\% of all voters live in counties with polling-place additions.

When voters experience polling-place assignment changes they might receive a mailer describing the new assignment. In some cases this mailer might serve as an advertisement for substitution voting methods, describing how voters can pursue alternatives. 
The effect that we see might in part be the result of substitution methods being advertised to voters. 
However, we do see a reduction in in-person voting across states with different access to substitutes, different forms of substitutions and likely different mailers.

\subsection{State-Supplied Polling-Place Location Files}
\label{app:states}
Public records requests were made to state election officials in all fifty states. In response to our requests, most states (\(31\)) provided partial records; the remaining states denied our request (\(1\)), failed to complete the request (\(10\)), did not have applicable records because of the prevalence of voting by mail (\(2\)), or were unable to fulfill the request because relevant polling place location information was only available at the county level (\(6\)). Of the states which provided partial records 18 were complete enough to be potentially included in the analysis, but the number of states with a sufficient number of voters varied by experimental design. Each address in the state-supplied files is geocoded, according to a process described in \secref{geocodes}.

For six states it is not possible to query a single state election official as polling-place locations are aggregated at the county level.
Instead, for the largest of these states (Texas) we 
 filed public records requests with county clerks in each of 254 counties. 
 We received partial records from 142 counties, but were only able to recover reliable polling-place location data in 106 of these.

\subsection{Filtering geocodes}
\label{geocodes}
State-supplied polling-place files describe each polling-place location with address descriptors such as street name, city, zip code and a textual place description. Our goal is to generate a \textit{geocode} for each location, using Google's geocoding API \footnote{\url{https://developers.google.com/maps/documentation/geocoding/overview}}. 
We form two potential addresses to geocode for each location. The first address is the full street address which consists of a street number, street name, street descriptor (such as st. or ave.), as well as the city and state. The second address is the place address which consists of the place description, city and state. For example, a polling place might be described as the Local Elementary School and the resulting place address might be Local Elementary School, Doeville TX. 

If the full street address is found to be precise and accurate according to the system below, its geocode is counted as valid and the associated latitude-longitude coordinates are used. 
If, however, the full street address fails either check, we instead attempt to geocode the place address. A geocode belonging to a place address must also be precise, but it only needs to pass the city and state accuracy tests.

\paragraph{Precision}
Each geocode is accompanied with metadata about the geocoding process. Here, we use a geocode's \textit{location type}. There are four potential location types: 
rooftop, geometric center, range interpolated and approximate. If a geocode's location type is classified as rooftop we accept it immediately, while we immediately reject all geocodes classified as approximate. 

If a geocode is classified with a location type of  geometric center or range interpolated we accept it conditionally. We would like to accept all location types of sufficient granularity; however some subjectivity is required to discern between categories of geometric centers or interpolated ranges. We take a conservative approach and retain geometric centers or interpolated ranges which are tagged as an establishment, store, local government office or point of interest. 

\paragraph{Accuracy}
Another item of metadata produced by the geocoding process is a formatted address of the geocode. 
We consider a geocode to be accurate if this formatted address matches the original address which was geocoded. 
To check accuracy we consider several components of an address, the street name, the street number, the city and the state. 

\begin{itemize}
\item{\textbf{City and State}}
For each city and state we check that the formatted city and state match the original city and state.
All geocodes whose formatted address city and state do not match the original city and state are considered inaccurate. 

\item{\textbf{Street Address}}
To check that the street names and numbers match we first parse a street address into components using the Python software package USADDRESS.\footnote{\url{https://github.com/datamade/usaddress\#readme}}
This package standardizes address components so that, for example, 1st would be transformed to first. However, upon inspection, this standardization 
is not universally applied. Consequently, we performed a second hand-coded round of standardization where all numerical street names and descriptors were encoded 
in long-form (i.e. first and street or avenue), respectively. Finally, we check that a geocode's formatted address street number and name match the originals. 

\item{\textbf{Evaluating Place Descriptions}}
If a street address is not considered to be precise and accurate, we instead geocode a location's place description. For example, a polling place might be described as Local Elementary School, Doeville TX. It is difficult to check the accuracy of a place description, as it will not contain an address, and descriptors returned from the geocoder might not match local descriptors. Instead of the full accuracy checks described above, for place descriptions we only check that the city and state of the place description's geocode match the original code. 
Thus, we only ensure that a geocode returned from a place description be precise, not that it be accurate. 
\end{itemize}

\subsection{Filtering to eligible registrants for analysis}
\label{app:filtering}
Beginning with potential voters from the TargetSmart data file we conduct a series 
of filtering steps to produce the datasets used for the shock and distance analyses. 
These steps are described at a high level in \tabref{tab:filter}
and in more detail in the following. 

Initially, we begin with a set of \textbf{potential voters in valid voting jurisdictions (counties and polling places)}. 
Within the states included in the analysis there are some counties 
which hosted vote centers in either 2012 or 2016. In these counties 
registrants can cast a ballot at any vote center regardless of their registration precinct. 
This environment is in stark contrast to that which we assume in these analyses
where voters must vote at their assigned precinct. 
Consequently, we remove all counties with vote centers in either 2012 or 2016 from 
the analysis.

\begin{table}
 \begin{center}
\resizebox{.89\columnwidth}{!}{%
\begin{tabular}{l r r}
\hline
\textbf{Filtering Step}& \multicolumn{2}{c}{\textbf{Number of potential voters remaining}}    \\
\hline
& \textbf{2012}& \textbf{2016}   \\
\hline
Potential voters in valid voting jurisdictions (counties and polling places) & 80,271,123 & 83,928,328 \\
Filter potential voters with valid address & 62,432,960 & 66,975,996\\
Filter potential voters with potential polling place assignment & 41,756,035 & 41,756,035\\
Filter to registered and plausible voters &  33,256,459 & 37,453,896\\
\begin{tabular}{@{}l@{}}Filter to registrants who live on a block where all pairs of registrants live \\  within .3 miles from one another\end{tabular}
 & 26,133,615 & 29,273,108\\
Filter to registrants who live on the same block in 2012 as in 2016 & \multicolumn{2}{c}{ 14,486,807}\\
Filter to registrants in \textbf{Distance analysis}: &\multicolumn{2}{c}{  249,309}\\
\hline
Filter to registrants in \textbf{Shock analysis}: &\multicolumn{2}{c}{  47,456}\\
\hline
\end{tabular}}
\caption{ Here, we detail the data filtering steps followed to create final data set. 
\label{tab:filter}}
\end{center}
\end{table}

\textbf{Filter to potential voters with valid address}
\label{sec:app_residences}
As addresses are used to ensure the quality of a polling-place assignment, and in \distance{} experiment, we require that voters have valid addresses in order to computer
their distance to the polling place. 
Several filtering steps are performed towards this end. 
We first drop all voters with missing address records. To compute a \unit{}'s distance to their assigned polling place we rely on geo-coordinates and we filter out all \units{}
with missing geo-coordinates. 

TargetSmart infers each voter's current address. We drop all voters whose current address does not match their registration address. 
Each geo-coordinate is accompanied with descriptors of the precision level at which it was recorded. For example, low precision levels are \textit{Extrapolate} or \textit{Zip Code} while a higher precision level is \textit{Street}. A geo-coordinate at the level of \textit{Extrapolate} refers to the closest known address to the original address and a  geo-coordinate at the level of \textit{Zip Code} refers to the centroid location of the zip code of the original address, while a geo-coordinate at the level of \textit{Street} 
refers to the street address of the original address. We retain only those addresses which are geocoded at the level of \textit{Street}.

Finally, each address is associated with a United States Postal Office (USPS) dwelling code. We retain only those residential addresses with a USPS code of High-rise, Building, or Apartment or Street Address. This removes all residences with a USPS code corresponding to a firm record, a general delivery area,  a Post Office box or a rural route or highway.

\textbf{Filter to potential registrants with potential polling place assignment}
\label{sec:app_infer}
To infer polling-place assignments for each voter we utilize both of our original data sources: the national voter file and the state-supplied polling-place files. 
Consider the three example voters in \tabref{table:example_voters} and the two example polling places in \tabref{table:example_places}.
Voter-1 and Voter-2 would be assigned to PP-1 while Voter-3 would be assigned to PP-2. We filter out all voters who are assigned to a polling place which we could not geocode. 

This matching process is vulnerable to errors  both in the original data sources and in the synchronization between them. 
We limit our analysis to \units{} whose inferred assignments are no more than 25 miles from their home  to remove egregious matching errors.

\commentout{
\begin{table}
 \begin{minipage}{.99\linewidth}
 \begin{center}
\resizebox{.99\columnwidth}{!}{%
\begin{tabular}{llllll}
\hline
\textbf{Voter ID}& \textbf{Address} & \textbf{City} &\textbf{State}& \textbf{Block ID}& \textbf{Precinct} \\
\hline
\hline
Voter-1 & 123 Main St& Milwaukee& WI& 1-Main-St-Milwaukee-WI &Cherry School 1 \\
Voter-2  & 125 Main St& Milwaukee& WI& 1-Main-St-Milwaukee-WI  &Cherry School 1  \\
Voter-3 & 2000 Third St& Milwaukee &WI& 20-Third-St-Milwaukee-WI &Apple School 1 \\
\hline
\end{tabular}

}
\caption{ Example voter records. 
}\label{table:example_voters2}
\end{center}
   \end{minipage}%
     \hfill\allowbreak%
      \begin{center}
    \begin{minipage}{.69\linewidth}
\resizebox{.99\columnwidth}{!}{%
\begin{tabular}{lllll}
\hline
\textbf{Place ID}&\textbf{Address}& \textbf{City} & \textbf{State} & \textbf{Precinct}  \\
\hline
\hline
PP-1 & 200 Main St & Milwaukee & WI & Cherry School 1\\
PP-2 & 1000 Third St & Milwaukee & WI & Apple School 1 \\
\hline
\end{tabular}
}
\caption{ Example polling place file. 
}\label{table:example_places2}
 \end{minipage}%
 \end{center}
\end{table}
}
\textbf{Filter to registered and plausible voters}
Next, we filter only to plausibly registered voters. 
That is we remove all unregistered voters and any voter who is marked 
as deceased. 
Hence, from this point on we refer to units in the analysis as registrants.

\textbf{Filter to registrants who live on a block where all pairs of registrants live within .3 miles from one another}
Our analysis operates on blocks of registrants, where each block can be divided into two faces (see \figref{assignment-shock} for an example). 
To identify eligible blocks of \units{} for our analysis, we create a block identifier for each voter.
This consists of all but the final two digits of the street number (e.g. 200 would be encoded as 2, 2100 would be encoded as 21), the street name, the street type, the city and state of a voter's residential address.
For example, the two \units{}, shown in \tabref{table:example_voters} to be residing at 123 Main St.\ and 125 Main St.\ in Milwaukee, Wisconsin, respectively, share
a block identifier.
 
To create a block identifier we require that an address consist only of numeric characters and that it be at least three digits long. The rational for the latter requirement is that in more rural locations where addresses are shorter there are examples of addresses with the same leading digit being on different blocks. For example, 80 Park Rd and 82 Park Rd might not be on the same block. We further ensure that all voters assigned to the same block live reasonably close by ensuring that no two voters with the same block identifier live more than 0.3 miles away from each other.

\textbf{Filter to registrants who live on the same block in 2012 as in 2016}
Finally, our analysis considers events that take place between the 2012 and 2016 elections. We therefore restrict our attention to registrants who resided on the same block in 2012 as in 2016. 

\textbf{Filter to registrants in analysis}
Finally, 
in both analyses we restrict our attention to blocks which meet the following
eligibility criteria. 
In the TargetSmart data
we observe instances where \units{} with the same registration address or the same registration geocode are assigned to different precincts.
We consider this to be an administrative error. One possible example of such an error is that a voter's registration is out-of-date and the record indicates
a previous polling-place assignment. To prevent these errors from effecting our analysis, we filter out all blocks where multiple polling
places are assigned to a single address or geocode. 
Additionally, we restrict our attention to those blocks where each block face 
has at least two voters.

Finally, for each analysis, 
we filter to all registrants still in the dataset who reside on a block
which is present in both the filtered 2012 and filtered 2016 data. 
For the shock analysis this produces 47,456 voters. 
For the distance analysis this produces 249,309 voters.

\end{document}